\documentstyle[12pt]{article}
\begin{document}

%\begin{flushright}
%cond-mat/01
%\end{flushright}
\vskip 1cm
%%%%%%%%%%%%%%%%%%%%%%%%%%%%%%%%%%%%%%%%%%%%%%%%
\begin{center} 
{\Large{\bf Lattice Dirac Fermions in a non-Abelian
Random Gauge Potential:
Many Flavors, Chiral Symmetry Restoration and Localization}}
\vskip 1cm

{\Large  Ikuo Ichinose\footnote{e-mail 
 address: ikuo@ks.kyy.nitech.ac.jp}}  
\vskip 0.5cm
{ Department of Electrical and Computer Engineering,
Nagoya Institute of Technology, Gokiso, Showa-ku,
Nagoya, 466-8555 Japan}

\end{center}

\vskip 0.5cm
\begin{center} 
\begin{bf}
Abstract
\end{bf}
\end{center}
In the previous paper we studied Dirac fermions in a 
non-Abelian random vector potential by using lattice 
supersymmetry.
By the lattice regularization, the system of disordered
Dirac fermions is defined without any ambiguities.
We showed there that at strong-disorder limit correlation function
of the fermion local density of states
decays algebraically at the band center.
In this paper, we shall reexamine the multi-flavor or 
multi-species case rather in detail
and argue that the correlator at the band center
decays {\em exponentially} for
the case of a {\em large} number of flavors.
This means that a delocalization-localization 
phase transition occurs as the number
of flavors is increased.
This discussion is supported by the recent numerical studies
on multi-flavor QCD at the strong-coupling limit, which
shows that the phase structure of QCD drastically
changes depending on the number of flavors.
The above behaviour of the correlator of the random Dirac
fermions is closely related with how the chiral symmetry 
is realized in QCD. 

\newpage

%%%%%%%%%%%%%%%%%%%%%%%%%%%%%%%%%%%%%%%%%%%%%%%%%%%%%%%%%%%%%%%%%
\setcounter{footnote}{0}
\section{Introduction}
In the last several years,
Dirac fermions in a random vector potential and/or with
random mass have been studied by various methods.
This system is closed related with the phase transition
between plateaus in the quantum Hall states and quasiexcitations
in the d-wave 
supercondctor\cite{ludwig,nersesyan,kogan,chamon1,
chamon2,castillo,caux,bhassen}.
However it is still controversial if there exist extended
states at the band center.
In the previous paper\cite{ichinose}, we studied the case of 
non-Abelian $U(N_c)$ random vector potential for large $N_c$.
There both weak and strong disorder cases are investigated
separately by using lattice supersymmetry(LSUSY) which 
was originally invented
for the study on the SUSY gauge theory.
Especially it was shown that correlation function of 
the fermion local density of states(DoS) decays
algebraically at the band center.
This result indicates the existence of extended states there. 

The above random Dirac fermion system on the lattice is
closely related with the lattice QCD for which there exist
many useful and definite studies by numerical simulations.
Among them, one of the most interesting studies shows that
the phase structure of lattice QCD changes drastically
depending on the number of flavors of quarks\cite{iwasaki}.
For small number of flavors, quarks are confined and 
the chiral symmetry is spontaneously broken, whereas
as the number of flavors is increased, quarks are deconfined
and the chiral symmetry is restored. 
Then mass of ``pions" is nonvanishing even for the vanishing bare quark mass
for many flavors.

The above result of multi-flavor
lattice QCD holds even at the strong-coupling limit which 
directly corresponds to the strong-disorder limit in the random
Dirac fermions.
In the previous study on the Dirac fermions in a random $U(N_c)$
vector potential\cite{ichinose}, we implicitly assumed a small 
number of fermion flavors.
In this paper we shall reexamine the system of multi-flavor
or multi-species Dirac fermions.
The flavor degrees of freedom is sometimes introduced in the
replica tricks and one takes the limit 
$N_f(=$the number of flavors)
$\rightarrow 0$ after calculation.
However in the present study, $N_f$ is kept finite and moreover
we even consider the case like $N_f/N_c >1$.
In the above lattice QCD, $N_c=3$ and the critical value
of $N_f$ is $7$.

Plan of the present paper is as follows.
In Sec.2, we shall study the strong-coupling limit of
lattice QCD by deriving an effective action of 
$U(N_c)$ invariant excitations, which is a kind of antiferromagnetic
spin model.
From this effective action, we can study the vacuum structure,
especially how the chiral symmetry is realized.
For a small number of flavors, the chiral symmetry is 
spontaneously broken and (quasi)massless pions appear
as a Nambu-Goldstone boson as we showed previously\cite{QCD}.
For a large number of flavors, on the other hand, we show
that the chiral symmetry is restored and ``pions" become
massive even at the chiral-invariant limit.
This result is of course in greement with the 
numerical studies\cite{iwasaki}.
In Sec.3, we review the lattice model of the Dirac fermions
in a $U(N_c)$ random vector potential.
Especially we explain the LSUSY which is
introduced in order to take ensemble average over the random vector
potential.
Then we shall derive an effective action by using similar
techniques used for QCD.
From the discussion on QCD in Sec.2, we show that there
is a phase transition as the number of flavors is increased.
This phase transition can be observed from the correlator
of the fermion local DoS.
Section 4 is devoted to conclusion.

%%%%%%%%%%%%%%%%%%%%%%%%%%%%%%%%%%%%%%%%%%%%%%%%%%%%%%%%%
\section{QCD with many flavors and chiral symmetry
restoration}
\setcounter{equation}{0}

Before going into detailes of the Dirac fermions in a random
vector potential, we shall investigate the strong-coupling 
limit of the lattice QCD.
We follow the disussion in Ref.\cite{QCD} which is also applicable
for the random fermion system rather 
straightforwardly\cite{ichinose}.

We consider $d$-dimensional hypercubic lattice.
Action of the QCD at the strong-coupling limit is given by
\begin{equation}
S_{QCD}={1\over 2}\sum \Big[\bar{\psi}(x)\gamma_\mu U_\mu(x)\psi(x+\mu)
-\bar{\psi}(x+\mu)\gamma_\mu U^{\dagger}_\mu(x)\psi(x)\Big],
\label{SQCD}
\end{equation}
where $x=(x_0,..,x_{d-1})$ denotes lattice site, 
$\mu=(0,..., d-1)$ is the direction index, 
$U_\mu(x)$ is $U(N_c)$ field on the link $(x,x+\mu)$ 
$U_\mu(x)=\Big(U_\mu(x)\Big)^a_b \in U(N_c)$ and we set 
the lattice spacing $a_L=1$.
The Dirac fermion $\psi$ carries not only 
the spinor and color indices $s=1,...,2^{[d/2]}$,  
$a=1,...,N_c$ but also flavor index $I=1,..., N_f$, i.e.,
$\psi=\psi_s^{a,I}$.
In Eq.(\ref{SQCD}), $\gamma_\mu$ is the $\gamma$-matrices
which work on the spinor index of $\psi$.
We add the following quark mass term in the action
\begin{equation}
S_M=-M_B\sum \bar{\psi}(x)\psi(x),
\label{mass}
\end{equation}
where $M_B$ is the quark bare mass.
The gauge field action is vanishing because we are considering
the strong-coupling limit.
In the numerical studies\cite{iwasaki}, the Wilson fermion is used for 
defining definite number of flavor.
Here we consider the naive lattice fermion.
However it is quite plausible that a similar phase transition occurs
as ``the number of flavor" $N_f$ is varied. 

It is well-known that $S_{QCD}$ in (\ref{SQCD}) can be
rewritten more compact form by the following change of
variables,
\begin{equation}
\psi(x)=T(x)\chi(x), \;\; \bar{\psi}(x)=\bar{\chi}(x)T^\dagger(x),
\label{transf}
\end{equation}
with $T(x)=(\gamma_0)^{x_0}\cdots(\gamma_{d-1})^{x_{d-1}}$ and 
using the identities like $(\gamma_\mu)^2=1$ and 
$(\gamma_0)^n\gamma_1=(-)^n\gamma_1(\gamma_0)^n$
($n$ is an integer), etc,
\begin{eqnarray}
S_{QCD}&=& {1\over 2}\sum \Big[\bar{\chi}(x)\eta_\mu(x)U_\mu(x)
\chi(x+\mu)-\bar{\chi}(x+\mu)\eta_\mu(x)
U^\dagger_\mu(x)\chi(x)\Big]  \nonumber   \\
&=&\sum\bar{\chi}(x)\hat{D}\chi(x),
\label{Schi}
\end{eqnarray}
where $\eta_0(x)=1, \; \eta_1(x)=(-)^{x_0}, \;
\eta_2(x)=(-)^{x_0+x_1}, \; \cdots$, and 
\begin{equation}
\hat{D}\chi(x)={1\over 2}\sum_\mu\Big[\eta_\mu(x)U_\mu(x)
\chi(x+\mu)-\eta_\mu(x-\mu)U^\dagger_\mu(x-\mu)\chi(x-\mu)\Big].
\label{hatD}
\end{equation}
Similarly
\begin{equation}
S_M=-M_B \sum \bar{\chi}(x)\chi(x).
\label{mass2}
\end{equation}
In Eqs.(\ref{Schi}) and (\ref{mass2}), the spinor indices
of $\chi$ and $\bar\chi$ are diagonal and they play the same role
as the flavor index.

Partition function of the system is given by 
\begin{equation}
Z_{QCD}=\int [DUD\bar\chi D\chi]e^{S_{QCD}+S_M},
\label{ZQCD}
\end{equation}
where $[DU]=\prod_{link}dU_\mu(x)$ is the Haar measure
of $U(N_c)$.
In Eq.(\ref{ZQCD}), the integral over the gauge field 
$U_\mu(x)$ can be performed by the $1/N_c$ expansion
because it reduces to the one-link integral in the 
strong-gauge-coupling limit\cite{QCD}.
There are two ``phases" in the one-link integral\cite{BG}.
The present case corresponds to the ``strong-coupling regime".
Detailes of the calculation can be seen in Ref.\cite{ichinose,QCD} 
and here we simply give the final expression.
To this end, we define the following color singlet
quark bilinear operators,
\begin{equation}
m^\alpha_\beta(x)={1\over N_c}\sum_a\chi^{a,\alpha}(x)
\bar{\chi}_{a,\beta}(x),
\label{mx}
\end{equation}
where $\alpha$ and $\beta$ are {\em flavor-spinor} indices
which take $N_{fs}=N_f\times 2^{[d/2]}$ values.
It is obvious that after integral over $U_\mu(x)$, the 
partition function $Z_{QCD}$ is expressed in terms of
$m^\alpha_\beta(x)$,
\begin{equation}
Z_{QCD}(J)=\int [D\bar\chi D\chi] e^{S_m(\lambda)+S_M+J\cdot m},
\label{ZQCD2}
\end{equation}
where $J$'s are source fields and $J\cdot m=\mbox{Tr}(Jm)
=\sum J^\alpha_\beta m^\beta_\alpha$.
In the leading-order 
of $1/N_c$\footnote{There is a systematic expansion in
powers of $1/N_c$. However, essential feature of the effective
action and the final results are not influenced by the
higher-order terms.}
\begin{equation}
{1\over N_c}S_m(\lambda)
=\sum_{x,\mu}\mbox{Tr}\Big[g(\lambda_\mu(x))
\Big], 
\label{Sm}
\end{equation}
with
\begin{eqnarray}
&& \lambda_\mu(x)=m(x)m(x+\mu),  \nonumber  \\
&& g(\lambda)=1-(1-\lambda)^{1/2}+\ln \Big[{1\over 2}
(1+(1-\lambda)^{1/2})\Big],
\end{eqnarray}
and Tr is the trace over the flavor-spinor 
indices.\footnote{We often omit the flavor-spinor indices
on $m(x)$ etc.}

The path-integral over $\chi$ and $\bar\chi$ in 
Eq.(\ref{ZQCD2}) can be performed by introducing
{\em elementary} meson fields ${\cal M}^\alpha_\beta(x)$.
In terms of ${\cal M}^\alpha_\beta(x)$, the partition
function is expressed as 
\begin{eqnarray}
Z_{QCD}(J)&=&\int [D{\cal M}] e^{S_{eff}({\cal M})
+J\cdot {\cal M}}, \nonumber \\
{1\over N_c}S_{eff}({\cal M})&=&\sum_{x,\mu}
\mbox{Tr}\Big[g(\hat\lambda_\mu(x))\Big]-\sum_x \Big[\mbox{Tr}
\ln {\cal M}(x)-M_B \mbox{Tr}{\cal M}(x)\Big],
\label{ZQCD3}
\end{eqnarray}
where $\hat\lambda_\mu(x)={\cal M}(x){\cal M}(x+\mu) $
and integral over 
${\cal M}$ is defined by polar decomposing ${\cal M}$ as 
${\cal M}=RV$ with positive-definite Hermitian matrix $R$
and $U(N_{fs})$ matrix $V$, and $\int d{\cal M}=\int dV$
($=$the $U(N_{fs})$ Haar measure)\cite{QCD}.

From the effective action (\ref{ZQCD3}), we can study the
vacuum structure of the system\cite{QCD}.
 ``Expectation value" of ${\cal M}(x)$
is obtained by the effective potential
which is obtained by substituting
${\cal M}^\alpha_\beta (x)=v\delta^\alpha_\beta$
in $S_{eff}$,
\begin{equation}
{1\over N_cN_{fs}} V_{eff}(v^2)=-d g(v^2)+{1\over 2}\ln v^2
-M_B v.
\label{Veff}
\end{equation}
From (\ref{Veff}), 
\begin{equation}
v=\sqrt{{2d-1 \over d^2}}+O(M_B).
\label{v}
\end{equation}

For vanishing bare quark mass $M_B=0$, $S_{eff}({\cal M})$
is a function of $\hat\lambda_\mu(x)$, and then
the vacuum expectation
value of the fields ${\cal M}(x)$ is given as 
\begin{equation}
\langle {\cal M}(x)\rangle =
\left\{
\begin{array}{rl}
vV_0 & \mbox{at even site}  \\
vV^\dagger_0 & \mbox{at odd site}, 
\end{array}
\right.
\label{vacuum}
\end{equation}
where $V_0 \in U(N_{fs})$.
From the above discussion, low-energy excitations 
are obtained by introducing ``pions" $\pi_K(x)$ as
\begin{equation}
{\cal M}(x) =
\left\{
\begin{array}{ll}
vV(x)=ve^{i\sum_K \pi_K(x) T^K} & \mbox{at even site}  \\
vV^\dagger(x)=ve^{-i\sum_K \pi_K(x) T^K}  & \mbox{at odd site}, 
\end{array}
\right.
\label{pion}
\end{equation}
where $T^K$'s are generators of $U(N_{fs})$. 

For $N_c/N_f > 1$, we can expect that fluctuation of $\pi_K(x)$
is not large as $S_{eff}({\cal M}) \propto N_c$
and topologically nontrivial configurations
do not contribute to the partition function.
Therefore we expand $V(x)$ as 
\begin{equation}
V(x)=1+i \sum_K \pi_K(x) T^K+\cdots, 
\label{expansion}
\end{equation}
and then
\begin{equation}
{1\over N_c} S_{eff}({\cal M})\sim -v^2\sum_{x,\mu}
\sum_K \Big(\nabla_\mu\pi_K(x)\Big)^2-M_B v \sum_x \pi^2_K(x).
\end{equation}
Interaction between the pions is suppressed by $1/N_c$.
In this case, the chiral symmetry is spontaneously broken
and $\pi_K(x)$'s are quasi-Nambu-Goldstone bosons\cite{QCD}.

What happends for $N_c/N_{fs} \ll 1$.
Numerical studies indicate that the ``pions" become massive
even in the limit $M_B \rightarrow 0$.
This means that chiral symmetry is restored in this case.
Actually this is not so difficult to see this behavior
in many-flavor case from $S_{eff}$.
The effective action (\ref{ZQCD3}) can be regarded as a
system of $U(N_{fs})$ antiferromagnets.\footnote{More
direct relationship between the lattice QCD at strong coupling
and antiferromagnets can be seen in the Hamiltonian
formalism\cite{smit}.} 
One can expect that
as degrees of freedom of the $U(N_{fs})$ spin, i.e., 
$V^\alpha_\beta$ in the present system, becomes
large, the long-range order is destroyed by fluctuations.
For example in spatial 2 dimensions, ordered phase exists 
in the Ising model whereas there is no ordered phase
in the $O(3)$ spin model.
Therefore we can naturally expect that the chiral symmetry
is restored for $N_{fs}/N_c \gg 1$.
Without the spontaneours breaking of the chiral symmetry, 
the expansion (\ref{expansion}), 
which assumes small fluctuation of $\pi_K(x)$,
is not justified.

Let us see the above chiral phase transition
more closely.
To this end, we shall introduce a more tractable model
which is expected to belong to the same universality
class of $S_{eff}$.
Let us notice that as $M_B \rightarrow 0$, 
$S_{eff}({\cal M})$ is a sum of the one-link term like
\begin{equation}
S_{link}=N_c \; \mbox{Tr}\Big[ g(\hat\lambda_\mu(x))-
{1\over 2d}\ln \hat\lambda_\mu(x)\Big].
\label{linkterm}
\end{equation}
Then we consider the following model which consists of a similar
one-link term, 
\begin{equation}
S_V={1\over g_0}\sum_{x,\mu}\mbox{Tr}\Big[
V^\dagger(x)V(x+\mu)+\mbox{H.c.}\Big],
\label{SV}
\end{equation}
where the parameter ${1\over g_0} \sim N_cv^2$.
Actually (a continuum version of) the action $S_V$ is derived directly 
from the 
QCD action (\ref{SQCD}) by using Zirnbauer's color-flavor 
transformation\cite{zir}(see also Ref.\cite{nishi}), and therefore 
it is quite plausible that $S_{eff}$ and $S_V$ belong to the same
universality class.

The system $S_V$ (\ref{SV}) can be easily studied by the 
mean-field theory(MFT).
To this end, let us assume the expectation value of 
the field $V(x)\in U(N_{fs})$ as 
\begin{equation}
\langle V^\alpha_\beta(x)\rangle =w\; \delta^\alpha_\beta,
\label{MFT}
\end{equation}
where we have used the $U(N_{fs})$ symmetry and we assume
that $w$ is real.\footnote{Please do not confuse $w$
with $v$. In the terminology of the spin model,
$v$ is the magnitude of spin, whereas $w$ is the magnetization.
Therefore nonvanishing $w$ means the long-range order.} 
We employ the following one-site action as MF action\cite{mft},
\begin{equation}
S_{MF}={w \over g_0}\mbox{Tr}\Big[V+V^\dagger\Big]
  -{1\over g_0}N_{fs}w^2,
\label{SMF}
\end{equation}   
and the partition function of the MFT is given by
\begin{eqnarray}
Z_{MF} &=& e^{-V_{MF}(w)} \nonumber  \\
  &=&\int dV\; e^{S_{MF}}.
\label{ZMF}
\end{eqnarray}
From Eqs.(\ref{SMF}) and (\ref{ZMF}), it is obvious that
the stationary condition of $V_{MF}(w)$ gives the ``gap"
equation which determines the expectation value $w$.

The integral over $V$ in (\ref{ZMF}) with the $U(N_{fs})$
Haar measure can be performed by the same techniques 
used before for integrating over the gauge field
$U_\mu(x)$,
\begin{equation}
V_{MF}(w) = -N_{fs}\; g\Big({4\over N_{fs}}({w\over g_0})^2\Big)
+{N_{fs}\over g_0}w^2.
\label{VMF1}
\end{equation}
Especially for small $w$, 
\begin{equation}
V_{MF}(w) = -\Bigg({w\over g_0}\Bigg)^2+{N_{fs}\over g_0}w^2
+O(w^4).
\label{VMF}
\end{equation}
From Eqs.(\ref{VMF1}) and (\ref{VMF}), it is obvious that for 
$N_{fs}>{1\over g_0}=N_cv^2$, $w=0$ is the stable minimum,
whereas for $N_{fs}<{1\over g_0}=N_cv^2$, $w=0$ becomes unstable
and $w$ has a finite value at the minimum of $V_{MF}$.
Then this simple consideration using $S_V$ and the MFT
supports our expectation, i.e., for $N_{fs}/N_c \ll 1$
the vacuum is chiral symmetric and there are 
no Nambu-Goldstone bosons,
whereas for $N_c/N_{fs}\ll 1$ the chiral symmetry is spontaneously
broken and gapless excitations exist.

In the chiral-symmetric case, physical quantities are
calculated by using the $U(N_{fs})$ Haar measure for
$V(x)$.
We are interested in the meson-meson correlation functions
like 
$\langle {\cal M}^\alpha_\beta (x) 
{\cal M}^\gamma_\omega(0)\rangle$.
The integral over $V(x)$ can be performed by using the
character-expansion for 
$V(x)V^\dagger (x+\mu)\in U(N_{fs})$\cite{ID}, and each
link term of the action can be written as;
\begin{equation}
\exp \Big(S_{link}\Big)
=\sum_\rho F_\rho (N_c,v)\chi_\rho(V^\dagger(x)V(x+\mu)),
\label{chara}
\end{equation}
where $\rho$ refers to irreducible representations of $U(N_{fs})$,
$\chi_\rho$ is the character and $F_\rho (N_c,v)$ is a
function of $N_c$ and $v$.

Integral over $V \in U(N_{fs})$ can be performed by
the following formula\cite{ID},
\begin{equation}
\int dV\;  V^{\dagger\alpha}_\beta V^\gamma_\omega
={1\over N_{fs}}\delta^\alpha_\omega\delta^\gamma_\beta.
\label{integral}
\end{equation}
From Eqs.(\ref{chara}) and (\ref{integral}),
the meson correlation function is calculated for the simplest
case, i.e., one-dimensional system with free boundary condition.
In this case the $V^\dagger V$ correlator is
obtained as follows;
\begin{eqnarray}
\langle V^{\dagger\alpha}_\beta (x) V^\gamma_\omega (0)
\rangle &\sim&  \Bigg({ F_f (N_c,v) \over N_{fs} F_0 (N_c, v)}
\Bigg)^{|x|}\; \delta^\alpha_\omega \; \delta^\gamma_\beta  \nonumber   \\
&=& e^{-m_\pi |x|}\;\delta^\alpha_\omega \; \delta^\gamma_\beta,
\label{mesoncor}
\end{eqnarray}
where $f$ and $0$ refer to the fundamental and trivial representations
of $U(N_{fs})$, respectively, and $m_\pi=\ln (N_{fs}F_0/F_f)$.
As the dimension of the group $U(N_{fs})$ becomes larger,
the mass $m_\pi$ is getting larger as we expected.

The above considerations naturally explain the numerical
calculations of the multi-flavor QCD\cite{iwasaki}.
In the following section, we shall apply the above observation
to the Dirac ferimons in a random vector potentail.
We shall begin with reviewing the previous studies using
LSUSY\cite{ichinose}.

%%%%%%%%%%%%%%%%%%%%%%%%%%%%%%%%%%%%%%%%%%%%%%%%%%%%%%%%%%%%%%

\section{LSUSY and localization}
\setcounter{equation}{0}
Let us start with reviewing lattice formulation of
random Dirac fermions.
Our model is motivated by the network model for the quantum
Hall phase transition in two-dimensional electron systems
in a strong magnetic field \cite{network}.
There electrons move along equipotential line of the random
impurity potential and acquire Aharonov-Bohm phase from
the strong magnetic field.
Then we shall consider $d=2$ case in this section.
Fermionic (electronic) part of the action is given as in the lattice QCD in 
the previous section;
\begin{equation}
S_D=\sum \bar\chi(x)\hat D\chi(x),
\label{SD}
\end{equation}
where $\chi(x)$ and $\bar\chi(x)$ are two-component Grassmann 
fields with the color $U(N_c)$ and flavor $U(N_{fs})$ indices 
as before.
The bare mass $M_B$ in Sec.2 corresponds to the energy measured from the
band center in the present electron system and we set $M_B=0$
as we consider states at the band center.
Corresponding to the Grassmann fields, we introduce complex
scalar fields $\phi(x)$ and $\phi^\dagger(x)$ with the color 
and flavor degrees of freedom as follows\cite{ichinose2},
\begin{equation}
S_\phi=\sum \hat D \phi^\dagger(x)\hat D \phi(x).
\label{Sphi}
\end{equation}
It can be shown that the system $S_D+S_\phi$ possesses
LSUSY,
\begin{eqnarray}
\delta\phi&=&\bar{\epsilon}_+\chi_-+\bar{\epsilon}_-\chi_+, \nonumber  \\
\delta\phi^\dagger&=&\bar{\chi}_-\epsilon_++\bar{\chi}_+
\epsilon_-,   \nonumber  \\
\delta\chi_{\pm}&=&-\hat{D}\phi\epsilon_{\pm},  \nonumber  \\
\delta \bar{\chi}_{\pm}&=&
-\hat{D}\phi^\dagger\bar{\epsilon}_\pm,
\label{SUSY}
\end{eqnarray}
where $\chi_{\pm}$ is the chiral decomposition, i.e.,
$\gamma_5 \chi_{\pm}=\pm \chi_{\pm}$ and
$\epsilon_\pm$ are anticommuting spinor variables
with chirality $\gamma_5\epsilon_\pm=\pm\epsilon_\pm$.
From the LSUSY (\ref{SUSY}),
the partition function is just unity, i.e.,
the determinant from the $\chi$ integral and that from
$\phi$ integral cancell with each other\cite{ichinose}.
Parameter $g$ which controls randomness of the vector potential
can be introduced as follows,
\begin{equation}
P[U]=\prod_x\exp \Big[{N_c \over g}\mbox{tr}
(U_\mu(x)+U^\dagger_\mu(x))\Big].
\end{equation}
Ensemble average with respect to the random vector potential
for calculating correlators etc can be performed readily
by the LSUSY\cite{ichinose};
\begin{eqnarray}
\langle {\cal O}\rangle &=& \int [DUD\bar{\chi}D\chi D\phi^\dagger D\phi]
P[U]e^{-S_D-S_\phi}{\cal O}  \nonumber  \\
&=& \int [DUD\bar{\chi}D\chi]P[U]
{e^{-S_D} \over [\mbox{det}(\hat{D}^\dagger \hat{D})]}\; {\cal O} \nonumber \\
&=& \int [DUD\bar{\chi}D\chi]P[U]
{e^{-S_D} \over \langle \chi;U|\chi;U\rangle}\; {\cal O},
\label{expec}
\end{eqnarray}
where ${\cal O}$ is physical observable of the Dirac fermion
and $|\chi;U\rangle$ is the lowest-energy state of $\chi(x)$ in the 
background of $U_\mu(x)$.
From Eq.(\ref{expec}), the utility of the LSUSY is obvious.
In the rest of the discussion,
we consider the strong-disorder limit $g\rightarrow \infty$ and
therefore there is no distribution weight for the gauge field $U_\mu(x)$
as in the strong-coupling limit of QCD.

The bosonic action $S_\phi$ can be rewritten into more compact 
form\cite{ichinose2,ichinose}, 
\begin{equation}
S_\phi\Rightarrow S_{\omega\varphi}
=\sum \Big[ \omega^\dagger \hat{D}\varphi+\varphi^\dagger
\hat{D}\omega\Big],
\label{Somega}
\end{equation}
where $\omega$ and $\varphi$ are complex boson fields.
Then the one-link
integral of $U_\mu(x)$ in the previous section can be 
applied straightforwardly.
Fermion-fermion(FF) sector is given by $S_m(\lambda)$ in 
Eq.(\ref{Sm}).
Boson-boson(BB) sector is also written in terms of the following
composite fields like
${1\over N_c}\sum_a \varphi^{a,\alpha}(x) 
\varphi^{\dagger}_{a,\beta}(x),\;
{1\over N_c}\sum_a \omega^{a,\alpha}(x) 
\omega^{\dagger}_{a,\beta}(x)$,
etc., but the sign of the action is just opposite to the
FF sector.
There are additional terms which include fermionic
composites like $\chi^a(x)\varphi^\dagger_a(x)$ but they 
contributions to the FF and BB sectors are negligible in the 
leading order of $1/N_c$.

From the above arguments and investigation on QCD in  
Sec.2, it is straightforward to
discuss physical properties of the random Dirac fields
at the band center which corresponds to the vanishing bare mass
$M_B\rightarrow 0$.
We are interested in the correlation function of the local fermion
DoS $m_\psi(x)=\sum_a\bar\psi_a(x)\psi^a(x)$, 
which is nothing but the ``pion" field in QCD.
Therefore in the case of a small number of flavors, 
one expects that the operator $m_\psi(x)$ condenses and has
a nonvanishing expectation value.
However as we are studying the $2$-dimensional system,
the genuine long-range order does not exist and the correlator of
$m_\psi(x)$ decays algebraically (quasi-long-range order) 
instead \cite{ichinose};\footnote{Here we do not show explicitly
the flavor-spinor indices as they are obvious as in 
(\ref{mesoncor}).}
\begin{equation}
\langle\bar\psi\psi(x)\bar\psi\psi(0)\rangle
\sim |x|^{-1/N_c}.
\label{corr1}
\end{equation}
The power-decaying behavior in (\ref{corr1}) comes from
the fluctuation of the ``pions" $\pi_K(x)$.
It should be remarked here that in the present lattice
formulation topologically nontrivial configurations of
the random vector potential are all included.
This formulation is motivated by the network model by
Chalker and Coddington\cite{network} and it is in contrast
to other formulations like that using the conformal field theory.
Therefore it is not so surprising even if the decay power of the
correlator (\ref{corr1}) is different from that obtained in other
methods\cite{bhassen}.
The power-decay behaviour (\ref{corr1}) means the existence of the extended
states anyway.

Let us turn to the many flavors case.
The discussion in Sec.2 can be applied straightforwardly
to the random system in this section.
As the system is $2$-dimensinal, 
the disordered phase is enhanced in the $U(N_{fs})$
antiferromagnets.
The correlator of the local fermion DoS
behaves as 
\begin{equation}
\langle\bar\psi\psi(x)\bar\psi\psi(0)\rangle
\sim e^{-M_l |x|},
\label{corr2}
\end{equation}
though we cannot estimate the ``inverse localization
length" $M_l$  analytically.
The above is a new and main result in this paper
which indicates that extended 
states at the band center disappear as the number of flavors
is increased.

Recently certain related model is studied in Ref.\cite{BRS},
and a similar phase transition is predicted, though the model is
defined in the continuum space instead of on the lattice.
In the present lattice model, which is motivated by the network model
for the quantum Hall transition, topological nontrivial configurations
of the gauge field are included.
Therefore it is quite nontrivial that a similar transition or crossover
occurs in the both cases.

%%%%%%%%%%%%%%%%%%%%%%%%%%%%%%%%%%%%%%%%%%%%%%%%%%%%%%%%%%%%%%%%

\section{Conclusion}
\setcounter{equation}{0}
In this paper we studied Dirac fermions in the random $U(N_c)$
vector potential by using lattice formulation.
Especially we showed that physical properties at the band
center drastically change as the number of flavors is increased.
For a small number of flavors, the correlator of the
local fermion DoS exhibits algebraic decay,
whereas for many flavors, it decays exponentially.
This result suggests that extended states at the band center
disappear as $N_f \rightarrow$large.
Numerical studies of QCD with many flavors\cite{iwasaki} supports
this result.

The above result is quite interesting and it is desired to
study the same problem by other methods.
Recently a closely related model, the random flux model, was 
studied by a lattice formulation\cite{altland}.
There a mathematical tool invented by Zirnbauer\cite{zir} was 
used and the DoS was calculated for the 
{\em single-flavor} case.
Then it is interesting to calculate the DoS for many flavors
and to see if qualitative difference appears as the number of
flavors is increased.
 
In the present studies we cannot estimate the critical value
of $N_f$ for the delocalization-localization transition(DLT).
Closely related and tractable model is a quantum $U(N_f)$
antiferrmagnetic Heisenberg model\cite{AFHM}.
There a phase transition, which corresponds to the DLT,
is predicted and the critical value of $N_f$ is estimated.

%%%%%%%%%%%%%%%%%%%%%%%%%%%%%%%%%%%%%%%%%%%%%%%%%%

\end{document}